\documentclass[preprint]{JHEP}
\usepackage{epsf}
\newcommand{\bea}{\begin{eqnarray}}
\newcommand{\be}{\begin{equation}}
\newcommand{\eea}{\end{eqnarray}}
\newcommand{\ee}{\end{equation}}
\newcommand{\nn}{\nonumber}

\title{Four-Fermion Effective Interactions and Recent Data at HERA}

\author{Nicola Di Bartolomeo$^{\S}$ and Marco Fabbrichesi$^{\dag \S}$\\
$^{\dag}$ INFN, Sezione di Trieste\\
$^{\S}$ Scuola Internazionale Superiore di Studi Avanzati\\
via Beirut 4, I-34013 Trieste, Italy.}

\abstract{We discuss the possibility of explaining the excess in
the cross section at high-$Q^2$ 
recently found at HERA in terms of four-fermion effective interactions.
To avoid the constraints from low-energy data, we select
three special cases in which the contact 
interaction is a product of vector-vector, axial-axial and
vector-axial currents. For these a satisfactory fit of HERA data is possible,
while keeping agreement with LEPII and CDF data, for an interaction
scale $\Lambda =3.5$ TeV.
As the statistics of the experiments improves---and if the effect 
persists---it will be  soon possible to discriminate between such
 contact terms and
the alternative explanation in terms of leptoquark exchange. }


\preprint{SISSA 34/97/EP \\
March 1997}

\begin{document}

\section {Introduction}

The excess in the neutral current cross section at large momentum 
transfered $Q^{2}$ recently announced at HERA~\cite{h1,zeus} allows us,
while we wait for an improvement in the statistics,
to speculate about physics beyond the standard model in a concrete
setting.

Actually, most of  
the territory is  already well mapped 
and  the theoretical 
analysis needed is available in the preliminary work done in view of the 
experimental runs at HERA~\cite{herabook} and in the older papers cited 
therein.

Such references convincingly argue that
any excess in the cross section similar to the one contained in the 
recent HERA data, can arise either
because of the exchange of a leptoquark, and has in this case the
shape of a resonance produced in the $s$-channel,  or because of 
the presence of an
effective four-fermion interaction, and has the shape of a continuum 
in the
$t$-channel instead.
In this letter we would like to investigate the latter 
possibility and then briefly compare the two.

That the presence of an effective four-fermion interaction
could lead, via interference with the Standard Model amplitude, to
an increase in the cross section was first pointed out in~\cite{eichten}. 
Such an interaction could originate from quark 
substructure or, more generally, by the exchange among quarks and 
leptons of a heavy particle 
like, for instance, an extra $Z'$ boson. Since the scale of such 
hypothetical new exchange is much higher than that probed at HERA, 
we
only see the contribution of  the effective four-fermion terms thus 
generated. Their form can in principle be rather complicated, as it is
the case for weak interactions in the Standard Model (an example of which
is the low-energy effective $\Delta S = 1$ lagrangian). 

As it is natural, as we were working on our paper, several works 
appeared discussing HERA data. Two of them are comprehensive 
analyses~\cite{alta,babu}, others are more specific to the leptoquark 
scenarios~\cite{CR,blum,kali,rizzo,DM}. 
Finally,  one work~\cite{BCHZ} is very close to our
approach and concentrates on four-fermion interactions;
 it reaches conclusions that, although not completely equivalent,  
 agree with ours.

\section{Low Energy Data and Contact Interactions}

In this section, we review the constraints on the four-fermion
couplings coming from the low-energy neutral current data. In particular,
we will see that atomic parity violation imposes
 quite severe bounds~\cite{ap}.

The most general helicity-conserving four-fermion contact interaction
among leptons and quarks is:
\be
{\cal L}^{4f} = \sum_{i,j =L,R} \eta_{ij}^{q}
\left( {\bar e}_i \gamma_\mu e_i )({\bar q}_j \gamma^\mu q_j \right)
\label{1}
\ee
It is conventional to write the couplings $\eta_{ij}^{q}$ as
\be
\eta_{ij}^{q} = \frac{ \pm 4 \pi}{(\Lambda_{ij}^{q})^2}
\label{2}
\ee

At HERA, the relevant operators are for $q = u,d$, and therefore there are
in general eight independent couplings  allowed. 

These eight couplings are reduced to six if we impose $SU(2)_L$ invariance,
giving the constraints:
\be
\eta_{LL}^{u} = \eta_{LL}^{d} \quad \mbox{and} \quad
 \eta_{RL}^{u} = \eta_{RL}^{d} 
\label{3}
\ee
It is important to remark that $SU(2)_L$ invariance gives a 
neutrino-quark four-fermion interaction, of the form
\be
{\cal L}^{\nu q} = ({\bar \nu}_L \gamma_\mu \nu_L) \left[
\eta_{LL} {\bar Q}_L \gamma^\mu Q_L +
\eta_{LR}^u {\bar u}_R \gamma^\mu u_R +
\eta_{LR}^d {\bar d}_R \gamma^\mu d_R \right]
\label{4}
\ee
where $Q_L$ is the $(u_L, d_L)$ doublet. The interaction (\ref{4}) can
produce effects in $\nu$-nucleon deep inelastic scattering, as discussed
in detail  below.

The low-energy parity-violating electron-quark interaction is usually
written by introducing  the coefficients $C_{iq}$, $i=1,2$, $q = u,d$, 
defined as
\bea
{\cal L}_{e q} & = & \frac{G_F}{\sqrt{2}} \left[{\bar e}\gamma_\mu \gamma_5 e \left(
C_{1u} {\bar u}\gamma^{\mu} u + C_{1d} {\bar d} \gamma^{\mu} d \right) +
\right. \nn \\
& + & \left. {\bar e}\gamma_\mu  e \left(
C_{2u} {\bar u}\gamma^{\mu} \gamma_5u + C_{2d} {\bar d} \gamma^{\mu} 
\gamma_5 d \right) \right]
\label{5}
\eea

The data on $C_{1q}$ expressed in terms of the so-called weak charge $Q_W$
\be
Q_W = -2 \left[ C_{1u} (2 Z + N) + C_{1d} (Z + 2 N ) \right]
\label{6}
\ee
which is measured with a precision of about $\sim 1.0\%$ in the isotope
133 of Cesium \cite{csnew}
\be
Q_W ( \mbox{Cs} ) = -72.11 \pm 0.27 \pm 0.89
\label{7}
\ee

Concerning the coefficients $C_{2q}$, only the combination 
\be
C_{2u} - \frac{1}{2} C_{2d} = -0.04 \pm 0.13 
\label{8}
\ee
is measured \cite{pdb}.

The four-fermion interaction  (\ref{1}) gives the following contributions
to the $C_{iq}$'s
\bea
\Delta C_{1q} & = & \frac{1}{2 \sqrt{2} G_F} \left ( -\eta_{LL}^{q} -
\eta_{LR}^{q} + \eta_{RL}^{q} + \eta_{RR}^{q}\right) \nn \\
\Delta C_{2q} & = & \frac{1}{2 \sqrt{2} G_F} \left ( -\eta_{LL}^{q} +
\eta_{LR}^{q} - \eta_{RL}^{q} + \eta_{RR}^{q} \right) 
\label{9}
\eea

As previously stated, the most stringent constraint 
on the $\eta_{ij}^{q}$ comes from the measure
of the weak charge (\ref{6}): assuming for simplicity $\eta_{ij}^{u} =
\eta_{ij}^d$, in absence of cancellation (i.e. turning on only one
$\eta_{ij}$) one has
\be
|\Delta Q_W (Cs) | \simeq \left( \frac{17 \; \mbox{TeV}}{\Lambda_{ij}} \right)^2 
\label{10}
\ee
Therefore  one easily gets a lower bound on $\Lambda$ 
of the order of 15 TeV, a scale too high to be probed at HERA.

To escape the bound (\ref{10}), we will select three  scenarios:
a vector-vector (VV) interaction, with
\be
\eta_{LL} = \eta_{RR} = \eta_{LR} = \eta_{RL} \equiv \eta_{VV}
\label{11}
\ee
an axial-axial (AA) interaction, with 
\be
\eta_{LL} = \eta_{RR} = -\eta_{LR} = -\eta_{RL} \equiv \eta_{AA}
\label{12}
\ee
and finally the product of a leptonic vector current with an 
axial quark current (VA), with
\be
-\eta_{LL} = \eta_{RR} = \eta_{LR} = -\eta_{RL} \equiv \eta_{VA}
\label{13}
\ee

The first two possibilities are parity-conserving, and therefore do not
contribute to the $C_i$: the last one (VA) contributes only to the 
$C_{2}$, not to the weak charge. One finds
\be
\Delta \left(C_{2u} - \frac{1}{2} C_{2d} \right) \simeq \left(
\frac{0.87 \;  \mbox{TeV}}
{\Lambda_{VA}}\right)^2
\label{14}
\ee
that means, see (\ref{8}), a lower bound on $\Lambda_{VA}$ of the order
of 2 TeV.

There also other possible combinations of couplings avoiding the 
atomic parity violation bounds, like for instance 
the one considered in \cite{BCHZ}
\be
\eta_{LR} = \eta_{LR} \quad \mbox{and} \quad \eta_{LL} = \eta_{RR} = 0
\ee
or a possible cancellation between the $u$ and $d$ contribution to
$Q_W$. To keep the discussion simpler, we shall restrict ourselves to
the three mentioned cases, and we will assume the same couplings to the $u$
and $d$ quarks. In \cite{nelson} it is noticed that  approximate global 
symmetries other than parity can eliminate the contribution of
contact terms to atomic parity violation.

If we require a $SU(2)_L$ invariant interaction, there is a neutrino-quark
four fermion term inducing possible deviations in 
 $\nu$-nucleon deep-inelastic scattering experiments. We stress however
that, being the measurements done with muon neutrinos, the data can constrain 
only a $(\nu_\mu \nu_\mu)(q q )$ ($q$ = $u,d$) contact term and therefore do
not strictly apply to our analysis. One has to take into account this constrain
if the interaction is family blind, as for instance in the case of an 
universally coupled heavy $Z'$. In such a situation 
the 
relevant lagrangian is, following the standard parameterization:
\bea
{\cal L}_{\nu q} & = & \frac{G_F}{\sqrt{2}}{\bar \nu} \gamma^{\mu} (1 - \gamma_5 )
\nu \nn \\
& \times & \sum_i \left[ \epsilon_L (i) 
{\bar q}^i \gamma_\mu (1-\gamma_5 ) q^{i} + 
\epsilon_R (i) {\bar q}^i \gamma_\mu (1+\gamma_5 ) q^{i} \right]
\label{15}
\eea
 The presence of the interaction (\ref{4}) will modify the standard
model values of the coefficients $\epsilon_{i}(q)$,
inducing the following deviations: 
\bea
\Delta \epsilon_L(q) & = & -\frac{1}{2 \sqrt{2} G_F} \eta_{LL} \nn \\
\Delta \epsilon_R(q) & = & -\frac{1}{2 \sqrt{2} G_F} \eta_{LR}^{q} 
\label{17}
\eea

Numerically, the expected deviation is of the order
\be
\Delta \epsilon_i (q) \simeq \left( 
\frac{0.62 \; \mbox{TeV}}{\Lambda}\right)^2,
\label{18}
\ee
which gives a lower bound on $\Lambda$ roughly of
the order of 4 TeV or larger~\cite{ccfr}.

\section{Four-Fermions Interactions at HERA}

We will now examine the effects of the interaction (\ref{1}) at the
colliders, and in particular we will try to see if it has the
potentiality to explain the HERA excess.
As explained in the previous section, we will limit our analysis to the
combinations of couplings (\ref{11}), (\ref{12}) and (\ref{13}).

We examine first the sensitivity of HERA to the different contact interactions
in the $e^+ p$ neutral current channel. In fig.\ 1 and 2, 
we plot the ratio of the differential
cross section $d \sigma/d Q^2$ including contact terms to the Standard Model
 differential cross section. The contact terms have a $\Lambda =
3.5$ TeV, and we assume the same contact terms for $u$ and $d$ quarks.
To avoid constraints from $\nu_\mu$ DIS and from the combined Tevatron data, 
we do not assume the same contact terms for muons.
The signs of the contact terms coefficients in fig.\ 1 are such to give the
maximum enhancement of the cross section, that is to interfere
constructively with the Standard Model contribution. This is obtained by
taking $\eta_{VV} = \eta_{VV}^+ > 0$, $\eta_{AA} = \eta_{AA}^- < 0$ 
and $\eta_{VA} = \eta_{VA}^+ > 0$ respectively in (\ref{11}), (\ref{12})
and (\ref{13}).
As one can see from the figure, the $VV$ contact interaction provides the
biggest effect, the $AA$ interaction is almost as big, and
the $VA$ contact interaction has approximately the same influence as the
$LR$ contact interactions, which is, among the chiral interaction 
$LL$, $LR$, $RL$ and $RR$, the one most constrained by HERA \cite{h195}. 
Of course the chiral interaction is excluded by the atomic parity violation
data, and we have put it in the figure only for
 comparison with previous analyses.

\FIGURE{              
\epsfxsize=12cm
\centerline{\epsfbox{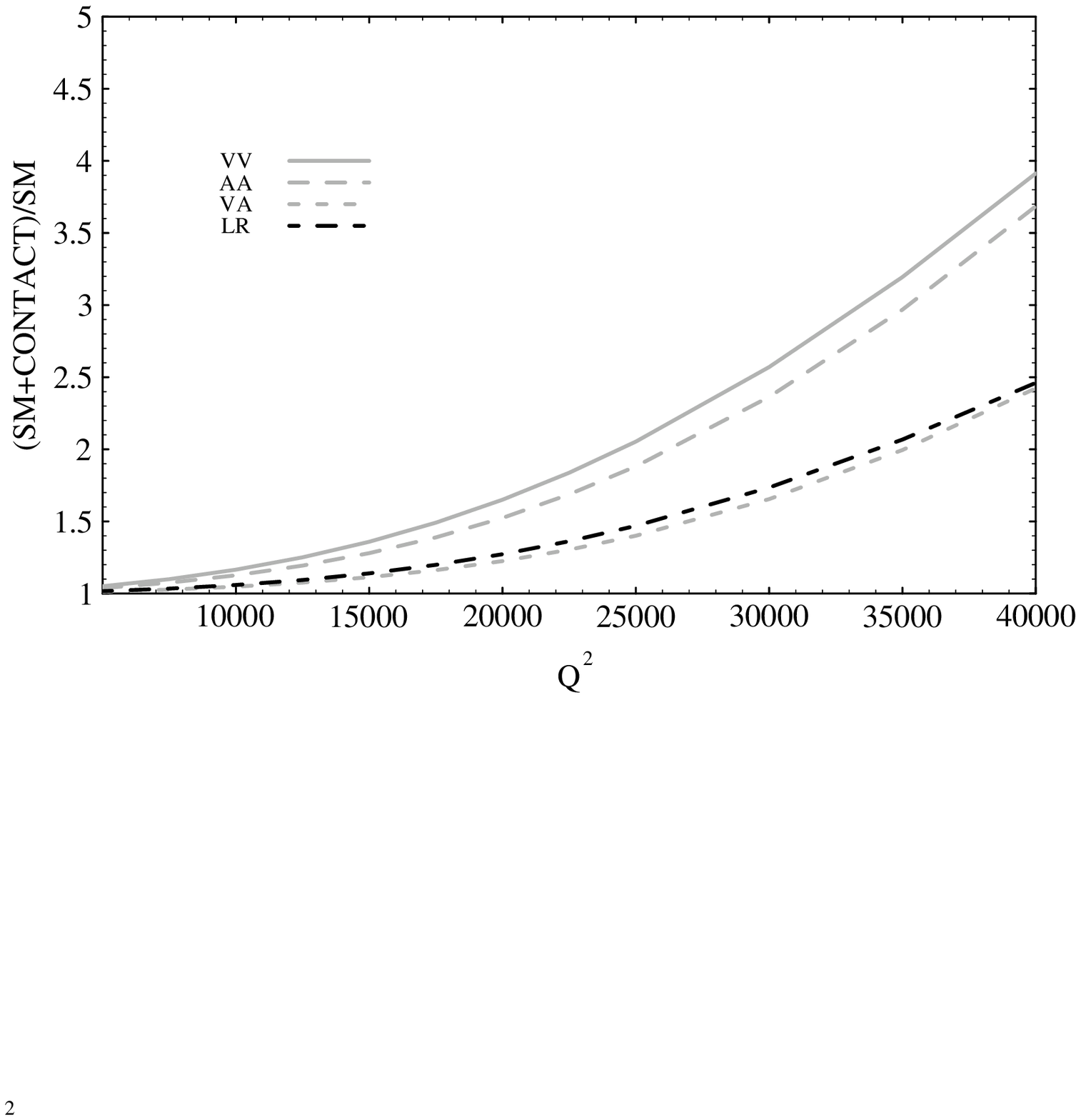}}
\caption{Effect of different contact interactions with
constructive interference with the Standard Model
on the
differential cross section $d \sigma/d Q^2$ as a function of $Q^2$
in GeV$^2$. The scale of the contact interactions is $\Lambda = 3.5$ TeV}}

The importance of the interference effect is evident in fig.\ 2, where
we have changed the signs of the contact term coefficients, plotting
as in fig.\ 1 the ratio of the differential
cross section $d \sigma/d Q^2$  to the Standard Model. 
Now the terms $\eta_{VV}^-$ and $\eta_{AA}^+$ reduce the Standard Model cross
section, while $\eta_{VA}^-$ is rather insensitive to the sign and
gives an excess comparable to $\eta_{VA}^+$. In other terms,
the interference is quantitatively very important for the $VV$ and
$AA$ cases, whereas it is negligible for
 the $VA$ contact term.

\FIGURE{              
\epsfxsize=12cm
\centerline{\epsfbox{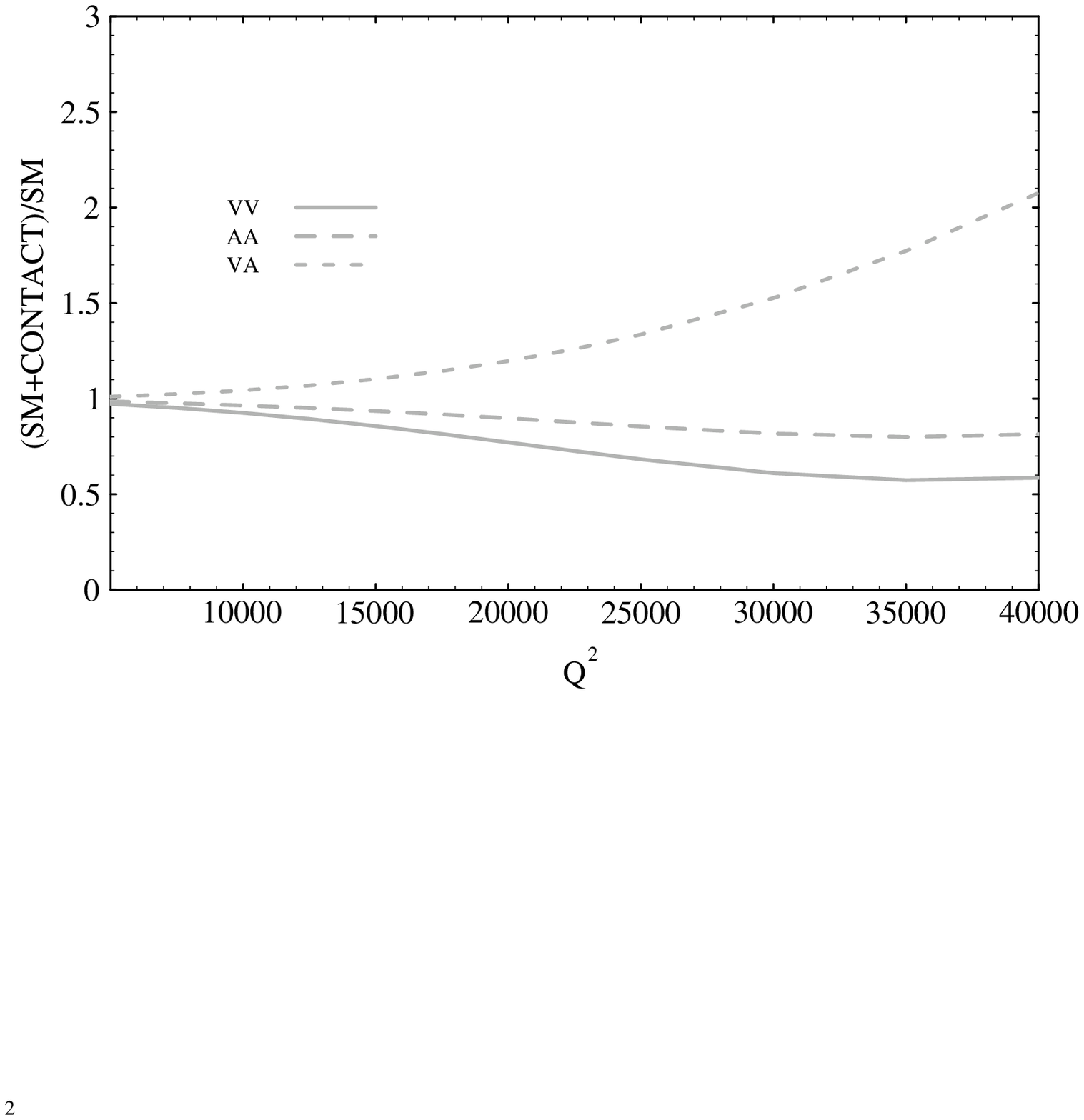}}
\caption{Effect of different contact interactions with
destructive  interference with the Standard Model on the
differential cross section $d \sigma/d Q^2$ as a function of $Q^2$
in GeV$^2$. The scale of the contact interactions is $\Lambda = 3.5$ TeV}}

Looking for an enhancement at HERA, we will obviously choose the
signs of the contact terms as those in fig.\ 1.

Beside HERA, the electron-quark interactions can potentially give
effects at CDF, in the  production of electron pairs, and at LEPII,
in the hadronic cross section. At LEP I, there is a one-loop 
potential effect on
the $Z^0$ leptonic width \cite{GN}.
As also noticed in  \cite{BCHZ}, the Standard Model amplitude changes sign 
under crossing from the $t$ channel $e q \to e q$ to the $s$ channel
$e^+ e^- \to q {\bar q}$ (LEPII) or $q {\bar q} \to e^+ e^-$ (CDF).
Therefore, by choosing the signs of the contact terms to
interfere constructively at HERA, we obtain destructive interference
at LEPII and CDF. 
This consideration helps to better understand the LEPII limits on the
contact interactions.

By looking at possible  deviations in the 
hadronic cross section, the OPAL collaboration \cite{opal} has obtained
the following limits
\bea
\Lambda_{VV}^- > 3.3 \; \mbox{TeV} \;& ; & \;\Lambda_{VV}^+ > 2.9 \;  
\mbox{TeV}
\nn \\
\Lambda_{AA}^- > 2.8 \; \mbox{TeV} \;& ; & \;\Lambda_{AA}^+ > 3.5 \; 
 \mbox{TeV}
\label{19}
\eea 

These bounds assume that the contact interaction is universal for all
five quarks produced at LEPII, and are weakened if only the first
family quarks $u, d$ are involved: in any case the ones concerning 
this analysis, $\Lambda_{VV}^+$ and $\Lambda_{AA}^-$, are below
the scale we will assume, that is $3.5$ TeV. 

The Drell-Yan production of $e^+ e^-$ pairs at CDF is another constraint
to be taken into account. As stated before, the chosen contact terms
will interfere destructively in this process.
A recent
analysis \cite{cdf} quotes the 95\% CL limits only on the left-left (LL)
contact interaction scale $\Lambda_{LL}$: $\Lambda_{LL}^- > 3.4$ TeV,
$\Lambda_{LL}^+ > 2.4$ TeV. We have checked that the interaction
$\Lambda_{VV}^+ = 3.5$ TeV , $\Lambda_{AA}^- = 3.5$ TeV and
$\Lambda_{VA}^+ = 3.5$ TeV
is still compatible with these CDF bounds, even if very close to the
exclusion region. We remark that by assuming the same contact scale for the
muons, i.e. a term $(\mu \mu)(q q)$ originating for instance from a heavy
$Z'$, the combined muon and electron CDF data give a stronger bound, 
excluding the value $\Lambda = 3.5 $ TeV for the selected interactions.

Having checked that HERA is very sensitive to the chosen forms of the
contact interaction, and that the chosen $\Lambda = 3.5$ TeV is
compatible with low-energy and colliders data,
we try now to see if the excess of high-$Q^2$ neutral
current
events reported at HERA could be explained by a contact interaction.

In fig.\ 3 we look at the $Q^2$ distribution of the events by 
integrating from a given minimum value $Q^{2}_{0}$. The standard 
model prediction, given by the solid black line,
is shown to lie below the combined data of H1 and 
ZEUS (the black dots and the gray error bars of the figure)
and the introduction of four-fermion interactions, represented by
the gray lines, is shown to help
in explaining most of the discrepancy.

Of course, the introduction of a leptoquark exchange would also
reproduce the behavior of experimental data, as shown in \cite{alta}.

In a way, fig.\ 3 is rather compelling  in pointing toward an actual
discrepancy between the standard model
 prediction and the data.

\FIGURE{              
\epsfxsize=12cm
\centerline{\epsfbox{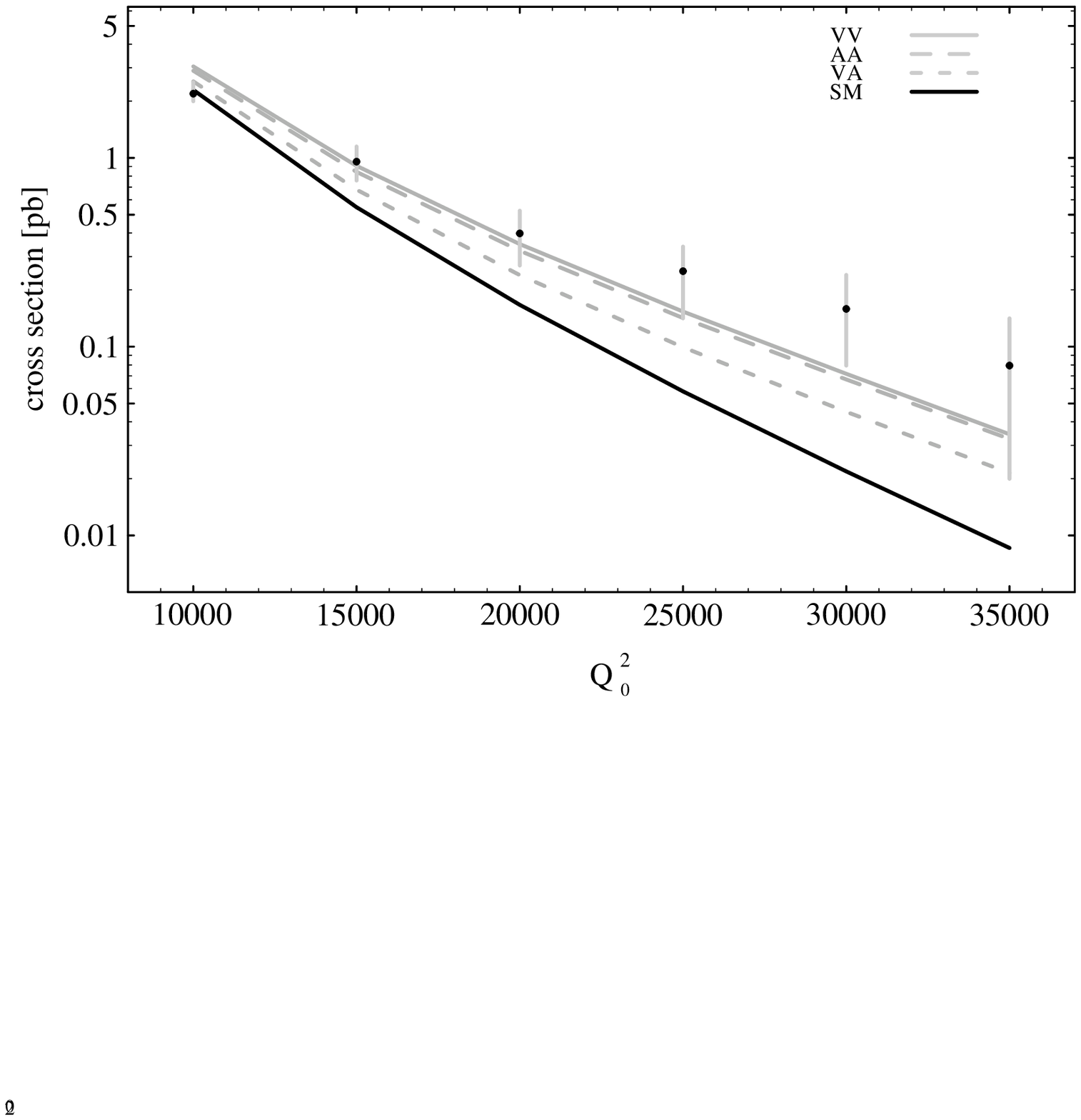}}
\caption{Integrated cross section for $Q^2 > Q^2_0$ versus 
$Q^{2}_{0}$ in GeV$^2$. The Standard Model (SM) prediction (the black line) is 
compared with the the contact interactions VV, AA and VA
(with $\Lambda = 3.5$ TeV) and with the
HERA data (H1 plus ZEUS).}}

Because of the integration over the values of $Q^{2}$, fig.\ 1 does not
helps us much in comparing  different scenarios, in particular
leptoquark versus contact interaction scenarios.
A more
discriminating way of plotting data and theoretical predictions 
consists in showing the cross sections versus the invariant mass.

To better understand such a plot, let us recall the kinematics of
deep inelastic scattering at HERA. In addition to the transferred 
momentum $Q^{2}$, there is another independent variable, that can be,
for instance the Bjorken variable
\be
x \equiv \frac{Q^2}{2 (P \cdot q)}
\ee
which itself leads to the invariant mass
\be
M = \sqrt{x s}
\ee
where $s$ is the center-of-mass energy squared, at HERA of about 
$(300 \: \mbox{GeV})^{2}$, or
\be
y = \frac{M^2}{x s} \, .
\ee

Fig.\ 4 shows the differential  cross sections 
$d \sigma/dM$, with the cuts $Q^2 > 15000$ GeV and $0.1 < y <0.9$,
for various scenarios. The lowest line (black) is the Standard 
Model prediction.
The VV and VA contact interactions with $\Lambda = 3.5$ TeV are 
the gray lines, respectively continuous and dashed. 
The AA line, being quite similar to the VV one, is not shown.
Moreover, we have plotted the cross section with exchange of a 
leptoquark  with mass of $200$ GeV as the dash-dotted line. In this case
of course one can see the resonance in the $s$-channel rising sharply at
$M = 200$ GeV. The leptoquark is a scalar $\phi$ coupled to the $d$ quark only 
\be
\lambda \phi {\bar d}_R e_L \; \; + h.c.
\label{20}
\ee 
with a value of coupling $\lambda = 0.04$, as indicated by the HERA
data \cite{alta,kali,rizzo}. This leptoquark is by convention named
as ${\tilde R}_{2L}$ \cite{buch} or ${\tilde S}_{1/2}$ \cite{camp},
and it is a very narrow state:
\be 
\Gamma = \frac{1}{16 \pi} \lambda^2 m_{LQ} = 6.37 \; MeV
\label{21}
\ee

Therefore, for  values of the invariant mass $M$ different from 200
MeV, the leptoquark scenario, as far as fig.\ 4 is concerned, 
is the same as the Standard Model. 
On the other hand,
the excess for the contact interaction is distributed over the
entire $M$ range.
  
\FIGURE{              
\epsfxsize=12cm
\centerline{\epsfbox{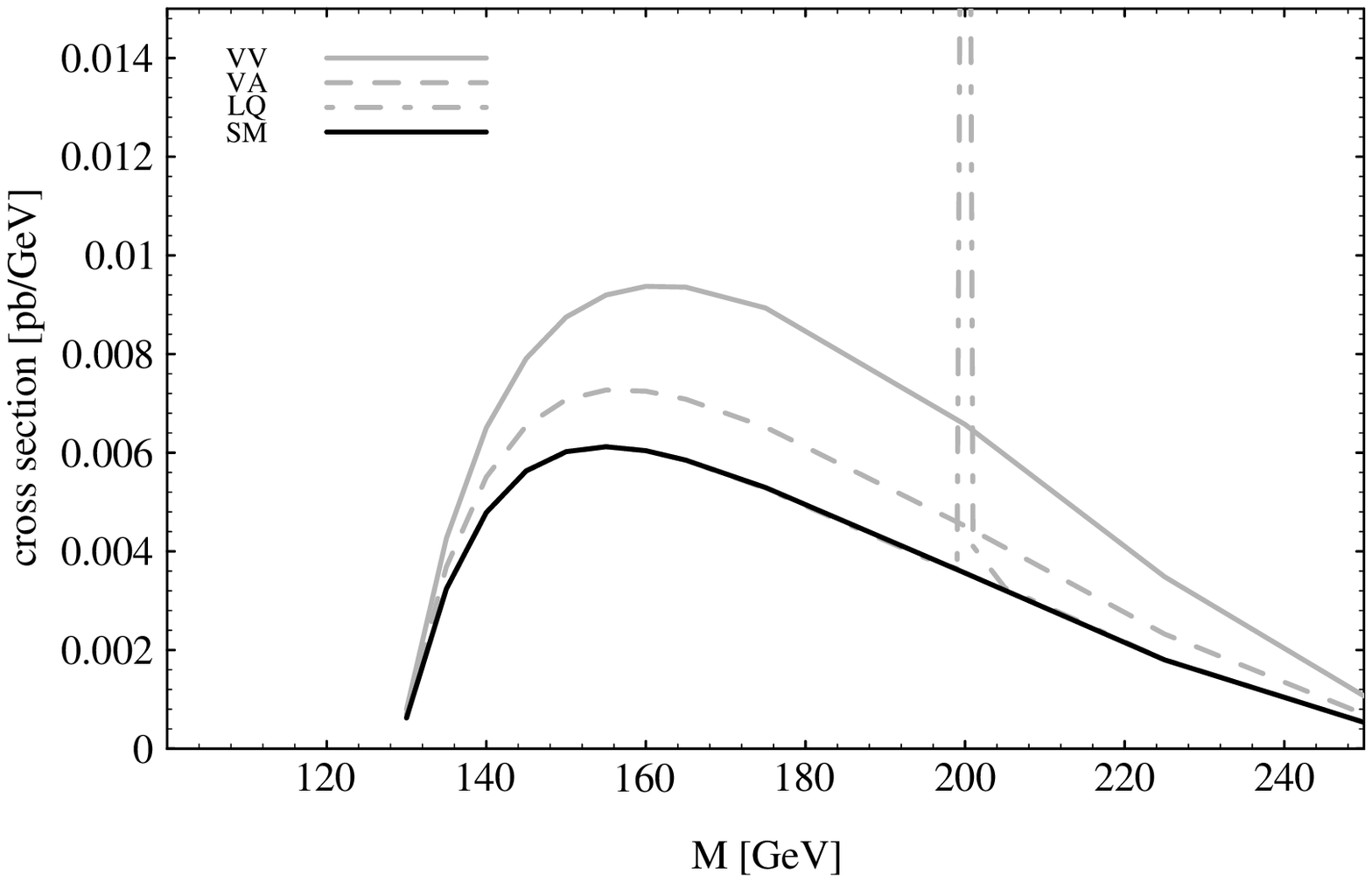}}
\caption{Differential cross section for various scenarios: contact
interactions VV e VA, with $\Lambda = 3.5$ TeV, and scalar leptoquark
(LQ), with coupling $\lambda = 0.04$.
The cuts $Q^2 > 15000 \; \mbox{GeV}^2 $ and $ 0.1 < y <0.9 $ are applied.}}

In order to compare with the experimental data, 
we have integrated the differential cross sections
of fig.\ 4 per bins of width $25$ GeV, computing the expected number of
events per bin at HERA. The histograms are presented in fig.\ 5 and 6.
As in fig.\ 4, there is a cut at $Q^2 > 15000 \; \mbox{GeV}^2$:
we have combined the H1 and ZEUS data, for a total of 24 events
passing the cut, taking into account of the efficiencies of
$\sim 80\%$. The $M$ distribution of the H1 
excess is peaked around $M = 200$ GeV, while the ZEUS data are
more evenly distributed. Moreover, there is also a discrepancy in the number
of events seen by the two experiments and 
it  has been argued \cite{drees} that the 
two experiments are compatible with a probability of less $1 \%$.

\FIGURE{              
\epsfxsize=12cm
\centerline{\epsfbox{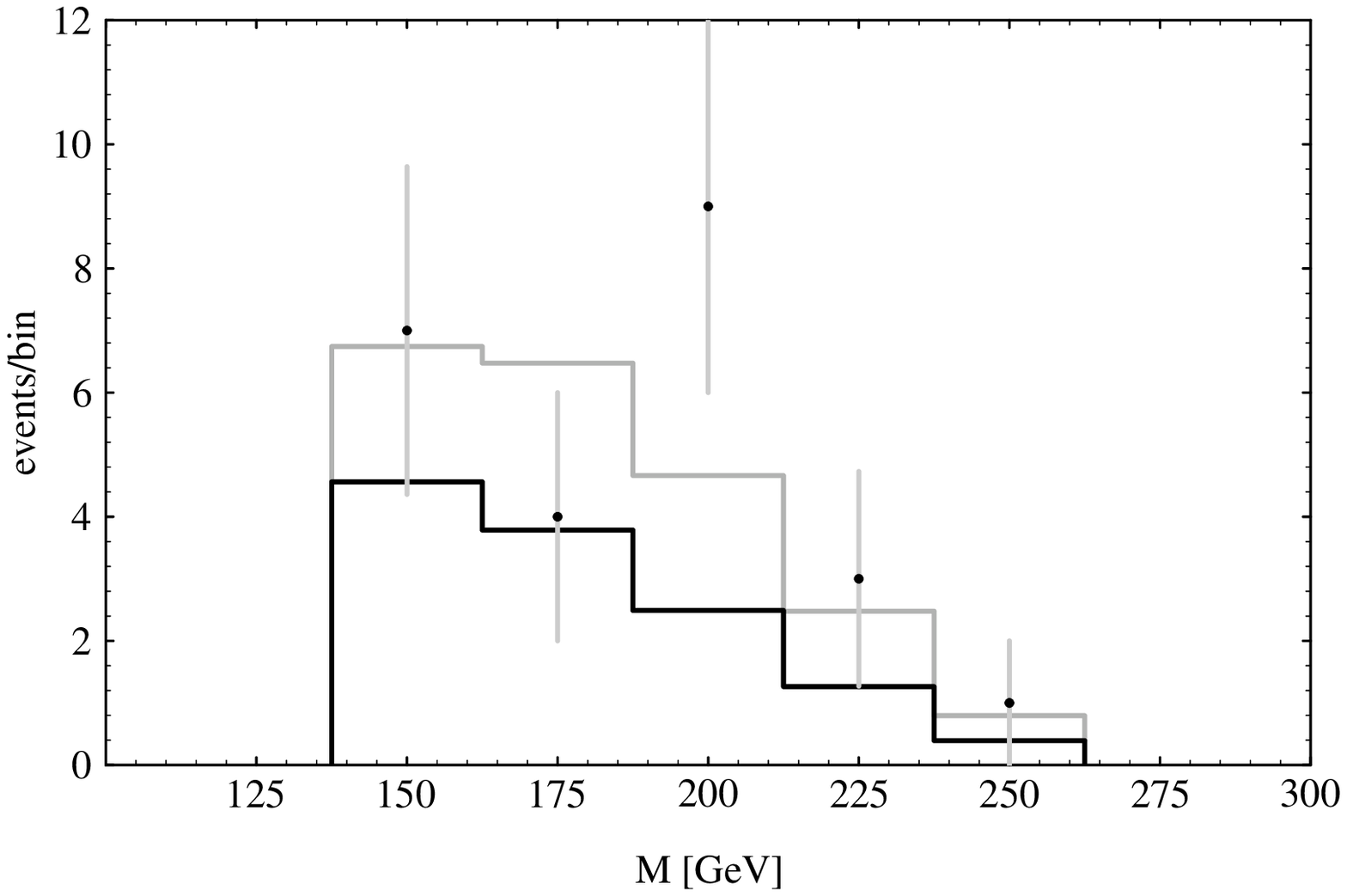}}
\caption{Events per bin vs. mass invariant: the VV four-fermion effective 
interaction scenario  (gray line) versus the Standard Model (black line), with
the cut $Q^2 > 15000 \; \mbox{GeV}^2$.}}

In fig.\ 5, we present the histogram of the expected events for a VV
contact interaction with $\Lambda = 3.5$ TeV (gray line),
together with the Standard Model expectation (black line).  
The VV contact interaction produces a distribution in  $M$ quite similar
in shape to the Standard Model, and accounts reasonably
well for the data. The only
discrepancy is seen in the bin centered at 200 GeV, because
of the seven H1 events clustered around that value of $M$.  

\FIGURE{              
\epsfxsize=12cm
\centerline{\epsfbox{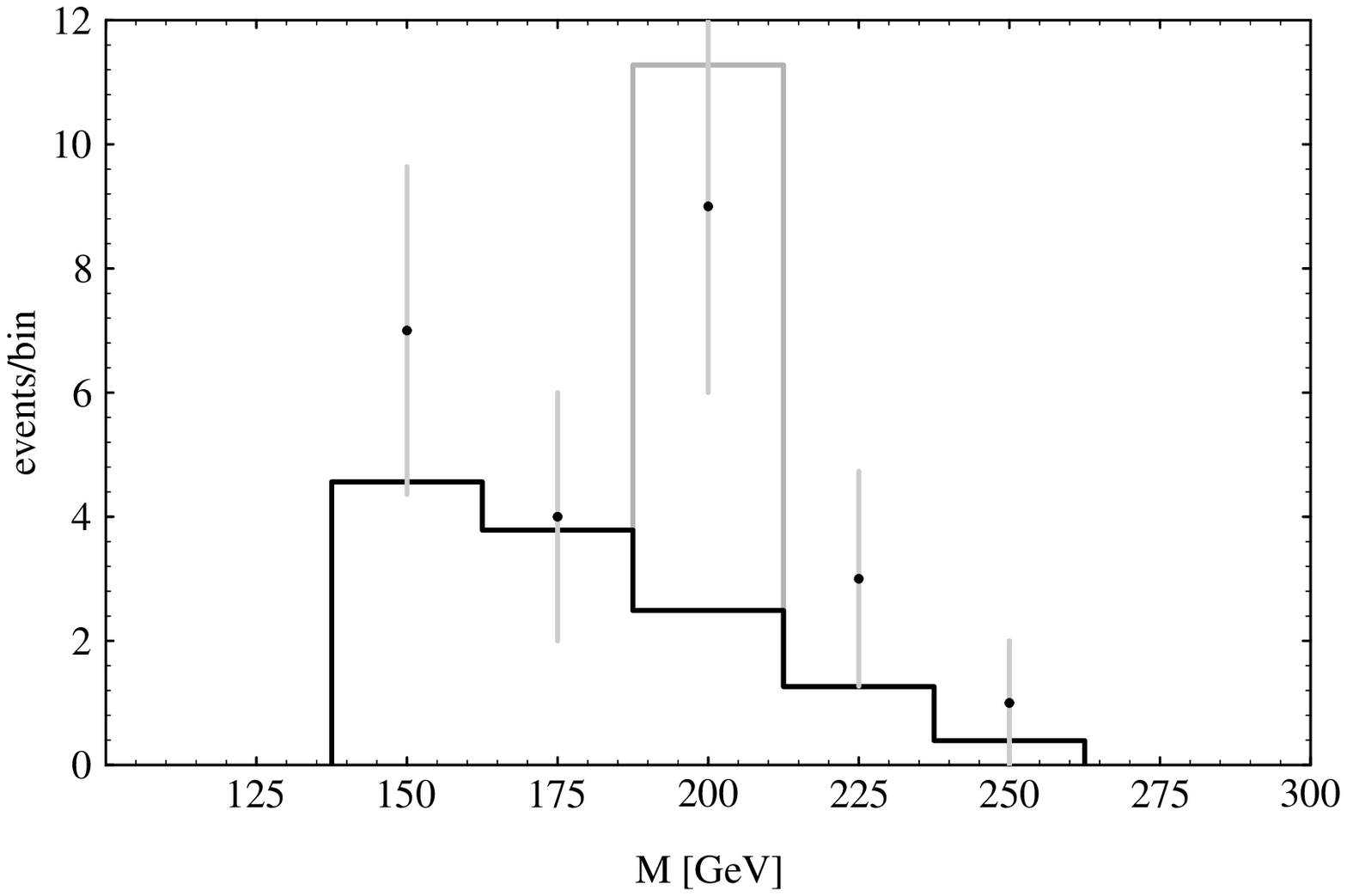}}
\caption{Events per bin vs. mass invariant: the leptoquark scenario
(gray line) versus the Standard Model (black line), with
the cut $Q^2 > 15000 \; \mbox{GeV}^2$.}}

Fig.\ 6 presents the leptoquark scenario, which provides an excess
only in the bin centered at $M = 200$ GeV, while agrees with
the Standard Model elsewhere. 

We think it is fair to say that the 
experimental situation does not exclude at the moment 
either the leptoquark or the contact interaction scenario, and that
only more data will clarify the situation. Only then we will also
know for sure whether we are in the presence of an actual discrepancy with
the Standard Model or a simple statistical fluctuation.

Forthcoming  analyses at Tevatron are expected to improve the
bounds both on leptoquark and contact interaction scenarios, and might even
exclude them. 
On the other hand the improvement in statistics at HERA
 will clarify the situation, as it can be readily understood by
looking at fig.\ 5 and 6. If the new data will be such to spread evenly in $M$
the excess, than the contact interaction scenario will be preferred, 
on the other hand, if the clustering at $M=200$ GeV will persist and 
be enhanced, then
it is clear that we will be looking at a leptoquark. 
Finally, if the new
data will
substantially reduce the number of events in excess, the Standard Model will
come out unchallenged one more time.

\acknowledgments{We would like to thank Ferruccio Feruglio 
for discussions and Ann Nelson for a relevant remark on neutrino contact
interactions.}

%
\clearpage
\renewcommand{\baselinestretch}{1}

\end{document}